\documentclass[twocolumn,switch]{article} 
\usepackage{lipsum}
\usepackage{tabularx}
\usepackage{balance}
\usepackage{xspace}
\usepackage{multirow}

\usepackage[ruled,vlined]{algorithm2e}
\newcommand{\etal}[1][]{%
\ifthenelse{\equal{#1}{}}{et~al.\xspace}{et~al.~\cite{#1}\xspace}%
}

\usepackage{listings}
\usepackage{xcolor}
\usepackage{enumitem}

\definecolor{keywordcolor}{rgb}{0.0, 0.0, 0.6}
\definecolor{commentcolor}{rgb}{0.0, 0.5, 0.0}
\definecolor{stringcolor}{rgb}{0.6, 0.0, 0.0}
\definecolor{backgroundcolor}{rgb}{0.95, 0.95, 0.95}

\lstdefinestyle{LangiumReqStyle}{
    backgroundcolor=\color{backgroundcolor},
    basicstyle=\ttfamily\footnotesize,
    keywordstyle=\color{keywordcolor},
    commentstyle=\color{commentcolor}\itshape,
    stringstyle=\color{stringcolor},
    numbers=left,
    numberstyle=\tiny,
    numbersep=5pt,
    showspaces=false,
    showstringspaces=false,
    showtabs=false,
    tabsize=1,
    captionpos=b,
    breaklines=true,
    frame=single,
    xleftmargin=8pt,
}

\newcommand{\citesec}[1]{Section~\ref{sec:#1}}
\newcommand{\citefig}[1]{Fig.~\ref{fig:#1}}

\newcommand{\ttformat}[1]{\texttt{\small {#1}}}
\newcommand{\hmreq}{\mbox{HM-Req}\xspace}
\newcommand{\hmreqcnl}{\mbox{HM-Req} \emph{CNL}\xspace}
\newcommand{\hmreqdash}{\mbox{HM-Req} \emph{Dashboard}\xspace}
\newcommand{\hmr}{HMR\xspace}
\newcommand{\hmrs}{HMRs\xspace}

\usepackage[mathlines, switch]{lineno} 

\usepackage[T1]{fontenc}
\usepackage{booktabs} 
\usepackage[hyphens]{url}
\usepackage{multirow}
\usepackage{rotating}
\usepackage{array}
\usepackage{color}
\usepackage{wasysym}
\usepackage{verbatim}
\usepackage{bigstrut}
\usepackage{multirow}
\usepackage{setspace}
\usepackage{tabularx}
\usepackage{url}
\usepackage{xspace}
\usepackage{pifont}
\usepackage{subcaption}
\usepackage[font=footnotesize]{caption}
\usepackage{array,booktabs}
\usepackage{soul}
\usepackage{vcell}
\usepackage{pifont}
\usepackage{afterpage}
\usepackage{rotating}
\usepackage{wrapfig}
\usepackage{caption}
\usepackage{graphicx}
\usepackage{caption}
\usepackage{enumitem}
\usepackage{lipsum}
\usepackage{balance} 
\usepackage[ruled,vlined]{algorithm2e}
\usepackage{algpseudocode}
\usepackage{tikz}
 
\newcolumntype{L}[1]{>{\raggedright\let\newline\\\arraybackslash}p{#1}} 
\newcolumntype{C}[1]{>{\centering\let\newline\\\arraybackslash}p{#1}} 
\newcolumntype{M}[1]{>{\centering\arraybackslash}m{#1}}
\newcolumntype{R}[1]{>{\raggedleft\let\newline\\\arraybackslash}p{#1}} 

\definecolor{alizarin}{rgb}{0.82, 0.1, 0.26}

\usepackage{multicol}
\usepackage{preprint}

\newcommand{\Description}[1]{}
\usepackage[numbers,square,sort]{natbib}
\usepackage[T1]{fontenc}

\usepackage{amsmath,amssymb,amsfonts}
\usepackage{graphicx}
\usepackage{textcomp} 
\usepackage{booktabs}
\usepackage[hyphens]{url}
\newcommand{\textcite}[1]{\cite{#1}}

\usepackage[utf8]{inputenc}	
\usepackage[T1]{fontenc}	
\usepackage{csquotes}
\usepackage{xcolor}		
\usepackage[colorlinks = true,
            linkcolor = purple,
            urlcolor  = blue,
            citecolor = cyan,
            anchorcolor = black]{hyperref}	
\usepackage{booktabs} 		
\usepackage{nicefrac}		
\usepackage{microtype}		
\usepackage{lineno}		
\usepackage{float}			

\usepackage{lipsum}		
\usepackage{orcidlink}

\usepackage{newfloat}
\DeclareFloatingEnvironment[name={Supplementary Figure}]{suppfigure}
\usepackage{sidecap}
\usepackage{longtable}
\usepackage{array}
\usepackage{tabularx}
\sidecaptionvpos{figure}{c}
\usepackage{titlesec}
\titlespacing*{\section}{0pt}{4pt plus 1pt minus 1pt}{1pt plus 0.5pt}
\titlespacing*{\subsection}{0pt}{3pt plus 1pt}{1pt}
\titlespacing*{\subsubsection}{0pt}{-5pt plus 1pt}{1pt} 
\usepackage{circledsteps}
\usepackage{tikz,xcolor,hyperref}
\usepackage{balance}

\title{HM-Req: A Framework for Embedding Values within  CPS Human Monitoring Requirements}

\usepackage{eso-pic} 

\AddToShipoutPictureBG*{%
  \ifnum\value{page}=1
    \put(0,20){%
      \makebox[\paperwidth]{\hfill \small\tt{\shortstack{Accepted for publication at the 34th IEEE International Requirements Engineering Conference (RE'26).\\*correspondence: Zoe.Pfister@uibk.ac.at}} \hfill}%
    }%
    
  \fi
}
\usepackage{authblk}

\author[1\thanks{\tt{Zoe.Pfister@uibk.ac.at}}]{Zoe Pfister \orcidlink{0009-0009-2882-5059}}
\author[1]{Ruth Breu}
\author[1]{Michael Vierhauser \orcidlink{0000-0003-2672-9230}}

\affil[1]{University of Innsbruck\\Department of Computer Science\\Austria}

\begin{document}

\twocolumn[ 
  \begin{@twocolumnfalse} 

\maketitle   

\begin{abstract}
Monitoring humans, for example, their movement or location, is essential for safe and efficient human-machine collaboration in Cyber-Physical Systems (CPS). This information allows CPS to ensure safety properties, adapt their behaviour dynamically, and coordinate with humans. To ensure that the design of a CPS respects ethical principles and the privacy of its stakeholders, system requirements, particularly those related to human monitoring, must reflect the human values of all involved stakeholders.
However, human values are often underrepresented in Software Engineering -- particularly during requirements elicitation and system design, crucial phases when introducing ethically critical functionality.
Stakeholder values are often implicit and conflicting, yet rarely systematically captured. Furthermore, unstructured natural language requirements introduce ambiguity and vagueness, complicating conflict resolution.
To address these problems, we propose HM-Req, a requirements elicitation framework including a Controlled Natural Language (CNL) for defining human monitoring requirements. These requirements are then augmented with human values from relevant stakeholders and integrated into a Value Dashboard to detect potential conflicts that require further discussion and resolution.
Validation results, applying the CNL to different datasets and conducting a survey and expert interview, provide evidence of the CNL's ability to capture diverse human monitoring requirements and demonstrate HM-Req's usefulness for requirements elicitation activities. 

\end{abstract}

\vspace{0.35cm}

  \end{@twocolumnfalse} 
] 

\section{Introduction}\label{sec:introduction}
Modern software-intensive systems increasingly operate in dynamic environments, where runtime monitoring is essential to ensure that these systems operate as intended and adhere to their requirements~\cite{zheng2016efficient,vierhauser2023runtime}. This is particularly relevant in the context of Cyber-Physical Systems (CPS), such as drones or robotic applications, where monitoring typically involves more than ``just'' collecting data from software or hardware components, but, more importantly, also humans interacting with the system in different ways~\cite{nikolakis2019cyber,gomez2025towards,cleland-huang_humanmachine_2023}.  
For example, in a human-machine collaborative assembly task, the machine needs to track the positions of humans in real-time to avoid collisions. Such systems may employ cameras for position tracking, or user-worn inertial sensors for gesture tracking~\cite{degeafernandez_multimodal_2017}.

\ieeecopyright

However, monitoring humans at runtime and collecting potentially sensitive information such as their location, movement, or activities poses significant challenges related to data protection, privacy, and other ethical aspects~\cite{grant1989monitoring}. 
In the aforementioned assembly task example, the sensors required to track an operator may be used to infer their performance, such as detecting if the operator frequently makes inefficient movements, or pauses the task.
In extreme cases, an operator may be deemed insufficiently productive, which could lead to termination of their contract~\cite{sartori2019monitoring}.
Whittle~\etal[whittlecase2021] argue for incorporating human values, such as privacy, security, or diversity, as first-class citizens in Software Engineering (SE). 
This means that software design activities, such as Requirements Engineering (RE), must integrate diverse human values during system design to avoid involuntary data sharing or misuse~\cite{detweiler2014value,perera2021impact}.
This is particularly true when requirements involve the monitoring of human attributes.
However, human values may differ substantially between stakeholders and across requirements, leading to the need for discussion and subsequent conflict resolution.
Stakeholders commonly articulate value-related concerns in natural language, which can be imprecise, ambiguous, and difficult to translate into actual requirements. 
At the same time, formal languages offer precision, but require specialized knowledge or experts when used, and are rarely accessible to non-technical stakeholders.
Value-based requirements engineering~\cite{ieee_computer_society_ieee_2021,whittlecase2021} addresses this need by connecting requirements to stakeholder values, providing mechanisms to evaluate trade-offs when conflicting goals arise. 
By designing a controlled natural language (CNL) for eliciting human monitoring requirements, we enable requirements engineers to formulate monitoring requirements using unambiguous vocabulary through a restricted set of available verbs and a well-defined structure, while preserving the natural language of the requirements. 
We define human monitoring requirements as \emph{\enquote{requirements that, to be implemented successfully, involve the continuous collection and processing of data about human stakeholder activities via one or more monitoring devices}}~\cite{Gil_2020,pfister_valuecomplemented_2025}.
These requirements can then be associated with stakeholder-value pairs, facilitating the automatic detection of potential value conflicts that must then be resolved through stakeholder discussions.

In this paper, we present the \textit{Human Monitoring Requirements framework} (\hmreq) that enables requirements engineers to specify human monitoring requirements in CPS and enrich them with stakeholder values derived from Schwartz's taxonomy~\cite{schwartz_universals_1992}.
This process is enabled through our \hmreqcnl that enforces a clear structure for defining human monitoring requirements. 
Additionally, we created the \hmreqdash, a proof-of-concept prototype that (1) allows mapping stakeholder values to human monitoring requirements and (2) automatically computes a \emph{Potential Value Conflict Score} between stakeholders based on Schwartz's smallest space analysis~\cite{schwartz_universals_1992}. The intention of our framework is not to replace established requirements modelling~\cite{gonccalves2025systematic,dalpiaz2016istar,fuxman2003formal} or value-based requirements engineering approaches~\cite{perera_continual_2020,thew_valuebased_2018}, but to complement them. \hmreq offers a distinct approach by providing a domain-specific CNL for defining human monitoring-focused requirements before associating them to stakeholder human values.
The contributions of this paper are as follows:

\begin{enumerate}[leftmargin=*]
    \item Grounded in existing datasets, we introduce a novel CNL for specifying human monitoring requirements in CPS. 
    \item We develop a method to associate the human values of each stakeholder with a human monitoring requirement and detect potential human value conflicts between them, implemented as a proof-of-concept prototype.
    \item We evaluate our framework by (i) assessing whether our CNL can capture real-world human monitoring requirements of diverse datasets, and (ii) by conducting a survey to collect information on the perceived usefulness of our \hmreqcnl and \hmreqdash.
\end{enumerate}
 
The remainder of the paper is laid out as follows. In \citesec{background}, we provide a brief introduction to domain-specific and controlled natural languages and present a motivating example. In \citesec{framework}, we introduce our \hmreq framework and its core components, and in \citesec{method-cnl}, we detail the process of creating \hmreqcnl.
We then introduce our research questions and validation in \citesec{results}. Finally, we discuss our findings in~\citesec{discussion}, threats to validity in~\citesec{threats}, related work in \citesec{relwork}, and conclude in \citesec{conclusion}.
\vspace{0.3em}

\section{Background and Motivating Example}\label{sec:background}

Requirements are commonly formulated using unstructured natural language (NL)~\cite{ambriola1997processing,gervasi2005reasoning}.
This, however, can lead to diverse issues such as word or phrase ambiguity, vagueness, or increased complexity~\cite{mavin_easy_2009,gervasi2005reasoning}.
Several approaches have been proposed in the RE community including template-based methods, such as EARS~\cite{mavin_easy_2009}, and MASTeR~\cite{kluge_schablonen_2024}, or goal-oriented methods~\cite{ali2010goal,elsood2014goal} that help formalize, structure, and prioritize requirements.
Template-based approaches further provide a clear sentence structure for requirements specification, aiming to address issues related to ambiguous and unclear requirements, but commonly do not restrict the vocabulary used, potentially leading to more ambiguity in the defined requirements~\cite{pohl_requirements_2025}. 
Particularly, when dealing with requirements about collecting potentially sensitive information, i.e., human monitoring requirements, it is important to also consider human values of relevant stakeholders~\cite{whittlecase2021,spiekermann_valuebased_2023,pfister_valuecomplemented_2025}. 

In the following, we discuss the background of (1) human values in RE, (2) domain-specific languages (DSL) and CNLs, and (3) WordNet and VerbNet, as the basis for the structured description of natural languages.

\textbf{Human Values and Requirements: }
To integrate human values into requirements specifications, a taxonomy of potential human values is required. A widely adopted taxonomy, including in SE research, is Schwartz's theory of basic human values~\cite{schwartz_universals_1992,schwartz_overview_2012}, which consists of 10 core motivational types of values and their respective sub-values that are recognized across cultures. Schwartz argues that the values form a circular structure, in which values adjacent to each other are complementary, while values with increasing distance between each other are opposing~\cite{schwartz_universals_1992}. 
For example, the universal value \emph{security} contains the sub-value national security, which is opposed to the universal value \emph{self-direction} that includes freedom.
We discuss related work of Values in SE in Section~\ref{sec:relwork}.

Integration of Human Values into the RE process has also received attention. In their Continual Value(s) Assessment framework, Perera et al.~\cite{perera_continual_2020} utilize the Schwartz's taxonomy, Feature Models, and Goal Models~\cite{pohl_requirements_2025} to relate requirements to specific values. Relationships can be either positive or negative, enabling evaluation of the degree to which a stakeholder's values are satisfied for a given combination of design choices from the Feature Model.
Supporting the elicitation of value-based requirements, Thew and Sutcliffe~\cite{thew_valuebased_2018} developed the VBRE method and validated it in industry. Their approach aims to guide the elicitation of stakeholders' values and motivations around sociopolitical concerns, while also accounting for stakeholders' potential emotional reactions to system change.

\textbf{Domain-Specific and Controlled Natural Languages: }
While conceptually similar, DSLs and CNLs are different regarding the level of formalism of the language. 
While DSLs are commonly more formal and used to create domain-specific programming languages~\cite{van_deursen_domain-specific_2000}, CNLs are constructed languages based on a natural language, aiming to preserve its NL properties while restricting their lexicon, syntax, or semantics~\cite{kuhn_survey_2014}.
We chose to develop a CNL in this work since preserving the NL properties of a requirement is beneficial for discussions with non-technical stakeholders~\cite{pohl_requirements_2025}. 

A CNL's natural language properties require grounding in existing languages.
We use both WordNet and VerbNet as a basis for the structure of our \hmreqcnl.
WordNet~\cite{princeton_university_about_2010} is a lexical database of English words \emph{\enquote{grouped into sets of cognitive synonyms (synsets) that each express a context}} and has been widely used in SE research, e.g., for semantic reasoning in NL queries~\cite{liu2020senet} and RE tasks~\cite{veizaga_systematically_2021}.
Each synset includes a definition and, in most cases, one or more brief sentences that demonstrate its use.
For example, the word \ttformat{monitor} is included in nine synsets, one of which (\ttformat{monitor.v.01}) includes the definition \textit{keep tabs on; keep an eye on; keep under surveillance} and the examples \enquote{we are monitoring the air quality} and \enquote{the police monitor the suspect's moves}.

VerbNet~\cite{university_of_colorado_boulder_verbnet_nodate} is a lexical resource of English verbs grouped into verb classes and mapped to WordNet, allowing retrieval of their corresponding synsets.
For example, verbs in the \ttformat{investigate-35.4} class (e.g., \ttformat{monitor}, \ttformat{surveil}, \ttformat{examine}) share a common syntactic structure: a noun phrase followed by a verb, a noun phrase specifying a location, and optional prepositional phrase (e.g., \emph{\enquote{The System monitors the environment [for workplace safety]}}).

To illustrate the advantages and obstacles of human monitoring at runtime in CPS, we present an example use case aiming to enhance worker safety in a shop-floor setting~\cite{pfister_valuecomplemented_2025}. 
In this scenario, the system must identify if a shop-floor worker enters a restricted or hazardous area, e.g., where autonomous robots operate. 
For this purpose, the system must continuously monitor a worker's location and alert them if they cross a designated boundary. Hence, a system that implements such a use case can enhance worker safety, but in turn may introduce privacy concerns. 
For instance, a manager may evaluate the performance of workers through monitoring their location and reprimanding them if they are deemed unproductive.
Ultimately, human values of the motivational types~\cite{schwartz_universals_1992} \textit{security} of a Product Owner (keeping workers \textit{healthy} by protecting them from threats), \textit{power} from a Manager (having \textit{authority} or \textit{preserving the public image}), and \textit{self-direction} of a Shop Floor Worker (\textit{freedom} / personal privacy) need to be traded off against each other.

\hmreq helps to clearly define requirements related to human monitoring and enrich these requirements with stakeholder human values for further value conflict analysis.

\section{Framework}\label{sec:framework}

To address the challenges of capturing requirements related to human monitoring and corresponding stakeholder values, and detecting potential value conflicts between them, we present our framework \emph{\hmreq} (cf.~\citefig{framework}). The goal, thereby, is twofold:
First, to help requirements engineers in defining unambiguous, structured human monitoring requirements (\emph{HMRs}) using a CNL, addressing problems that arise in natural language requirements~\cite{mavin_easy_2009}. 
Second, to enable the enrichment of these requirements with human values~\cite{schwartz_universals_1992,pfister_valuecomplemented_2025} to detect potential value conflicts and provide a basis for conflict resolution in stakeholder meetings.
We developed \hmreq following a Design Science approach~\cite{hevnerDesignScienceInformation2004}.

\begin{figure}[t!]
    \centering
    \includegraphics[width=0.90\columnwidth]{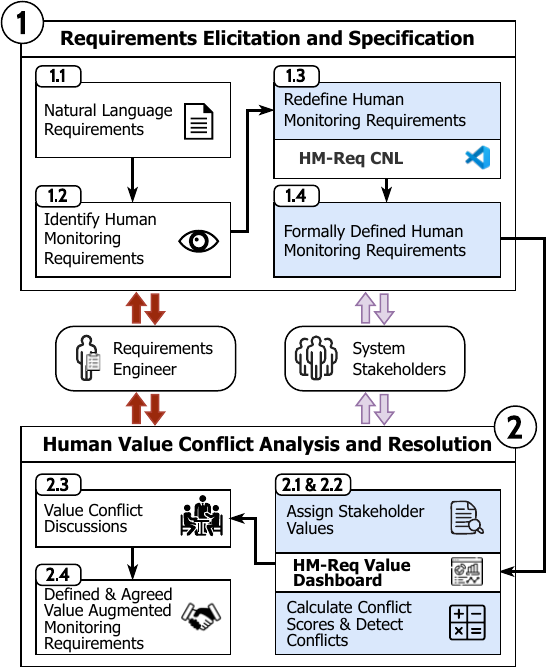}
    \caption{Our Proposed \hmreq Framework, consisting of Requirements Elicitation and Specification, and Human Value Conflict Analysis and Resolution.}
    \label{fig:framework}
    \vspace{-1em}
    \Description{Our Proposed \hmreq Framework, consisting of Requirements Elicitation and Specification, and Human Value Conflict Analysis and Resolution.}
\end{figure}

\subsection{Elicitation \& Specification of Requirements  with \hmreq}
Before \hmrs can be specified, they first need to be identified or derived from existing functional and non-functional requirements specifications. This starts with (1.1) existing system requirements, e.g., specified in natural language in a requirements document. 
Then, (1.2) domain experts are required to help identify \hmrs from the retrieved set of requirements. This initial step involves manually filtering for requirements that involve the collection and processing of data about human stakeholder activities. 
In some cases, the requirements already provide some assumptions and hints on how the monitoring should be implemented (e.g., using GPS). 
The outcome of this step is a set of \hmrs specified in unstructured natural language, which may still be ambiguous in their vocabulary and vague in definition.
When considering our motivating example (cf.~\citesec{background}), an unstructured \hmr could be \emph{\enquote{The location of Shop Floor Workers shall be tracked (GPS) while they are working in dangerous areas.}}.

While this example already contains relevant information, it does not conform to a clear syntactic structure nor provide a list of all involved stakeholders required for completion, which is necessary for later stakeholder value assignments and discussions.
To solve this, we redefine each \hmr using our \hmreqcnl (1.3). 
The \hmreqcnl draws inspiration from EARS~\cite{mavin_easy_2009}, a structured template for writing clear and consistent software requirements, and the work of Veizaga et al.~\textcite{veizaga_systematically_2021}. 
We further limit the available syntactic structures and vocabulary, aiming to reduce common problems that arise in natural language requirements, such as ambiguity, through a predefined set of usable verbs in the requirement block, or complexity, by enforcing a predefined structure~\cite{pohl_requirements_2025,mavin_easy_2009}.


\textbf{CNL Creation:} To create the \hmreqcnl, and establish its grammar, we selected \hmrs from five openly accessible requirements datasets~\cite{ACTIVAGE_dataset,MobSTr_dataset,PROMISE_exp_dataset,VHCURES_dataset,who_dataset} and analysed their structure and vocabulary through natural language processing.
Inspired by the approach of Veizaga et al.~\textcite{veizaga_systematically_2021}, we extract the verb lemmas\footnote{A lemma is a word's base form (e.g., the lemma of \emph{monitoring} is \emph{monitor}).} of each requirement.
For each extracted verb, we then retrieve its respective VerbNet class and add all possible senses of the lemma.
For each VerbNet code we found in this process, we leveraged the syntax definitions found in VerbNet 
to create the basic structure of the grammar. This resulted in the implementation of a total of 51 VerbNet class-based grammar rules with a total of 87 verbs.
An example of a grammar rule based on the VerbNet class \texttt{advise-37.9} can be seen in Listing~\ref{lst:advise-37-9}. We go into further detail on the deduction of the \hmreqcnl in~\citesec{method-cnl}.


Before defining requirements using our \hmreqcnl, a requirements engineer must define the relevant  stakeholders and actors. 
Listing~\ref{lst:requirement_structure} provides an overview of the structure of a requirement defined using our CNL. 
Each requirement starts with the keyword \ttformat{req}, followed  by a unique identifier. The first part of the requirement is similar to EARS, starting with preconditions or triggers (\ttformat{EARSPreStatement} rule), followed by the main requirement content. This part consists of (1) the actor or system name, (2) the modal verb (e.g., \emph{shall}), (3) the requirement block -- with  custom grammar rules extracted from each selected VerbNet class of our requirements -- and (4) a list of stakeholders that are relevant for completion of this requirement. 
Relevant stakeholders are stakeholders that may be affected by the monitoring aspect of the requirement or need to be involved in requirements discussions.
The outcome of redefining requirements using our \hmreqcnl is a set of structured \hmrs, each including its relevant stakeholders.

\begin{figure}[]
\begin{lstlisting}[caption={Example of the Grammar Structure of \texttt{advise-37.9}.}, label={lst:advise-37-9}]
ADVISE_37_9:
    verb=('notify' | 'alert' | 'inform')
    ArticleInLowercase? recipient=[Actor]
    possibleRestriction=('about' | 'of' | ...)?
    topic=STRING?;
\end{lstlisting}
\vspace{-1.5em}
\end{figure}
\begin{figure}[]
\begin{lstlisting}[caption={High-Level Structure of the \hmreqcnl Grammar.},label={lst:requirement_structure}]
Requirement:
    'req' requirementID=ID ':'
    earsPreStatement=EARSPreStatement
    requirementContent=RequirementContent
    relevantStakeholders=RelevantStakeholders;

RequirementContent:
    actor=[Actor] ModalVerb Not? 
    requirementBlock=RequirementBlock '.';
\end{lstlisting}
\Description{High-Level Structure of the \hmreqcnl Grammar.}
\vspace{-2em}
\end{figure}

\citefig{grammar-example} shows how the previously unstructured requirement of our motivating example is redefined using the \hmreqcnl. 
The example uses the structure defined in the grammar rule \emph{advise-37.9} (cf. Listing~\ref{lst:advise-37-9}). 
These rules and the respective names originate from their corresponding VerbNet class. 
In this example, the rule allows the use of the keywords \emph{notify}, \emph{alert}, or \emph{inform} followed by an actor (\emph{the Shop\_Floor\_Worker}) and ending with an optional restriction keyword (i.e., \emph{about}) and topic (i.e., \emph{"leaving the area"}).
The requirement ends with a listing of all relevant stakeholders.
In the example, the \emph{Shop\_Floor\_Worker}, the \emph{Manager}, and the \emph{Product\_Owner} are relevant, as the \textit{Worker} will be monitored, the \textit{Manager} may be overseeing the location of \textit{Workers}, and the \emph{Product\_Owner} is concerned about the safety of their workers.

\subsection{Value Conflict Analysis and Resolution}
\label{sec:value-assignment-conflict-resolution}
Once requirements are structured using our \hmreqcnl, the next step is to enrich them with the human value of each relevant stakeholder, enabling systematic definition of stakeholder values and identification of potential conflicts among them (e.g., freedom vs. authority). 
This step addresses a critical gap identified by Whittle~\etal[whittlecase2021], who argued that \enquote{human values are heavily underrepresented in SE methods}. 
The assignable human values are based on the original 56 human values presented by Schwartz~\textcite{schwartz_universals_1992} (e.g., \emph{power - authority} or \emph{self-direction - freedom}) and also include the reasoning of why a stakeholder chose a specific value for a requirement, articulated as a short \emph{value statement}.
We chose the value framework by Schwartz since it is widely used in existing SE literature~\cite{whittlecase2021,wohlrab_supporting_2024,ferrario_applying_2023} and, with its smallest space analysis (SSA)~\cite{guttman_general_1968}, provides clear relationships between human values in 2D space.
While there exist methods to elicit values during the RE process~\cite{thew_valuebased_2018} or software design~\cite{friedman_value_2013}, they do not explicitly use the Schwartz value framework. We assume values to be elicited together with stakeholders during workshops.
Once values are captured, a requirements engineer, in collaboration with stakeholders, can then manually go through each requirement and try to find potential value conflicts between stakeholders. However, manually detecting such conflicts, particularly involving numerous requirements, is a non-trivial and often tedious task.

\begin{figure}[t]
    \centering
    \includegraphics[width=\linewidth]{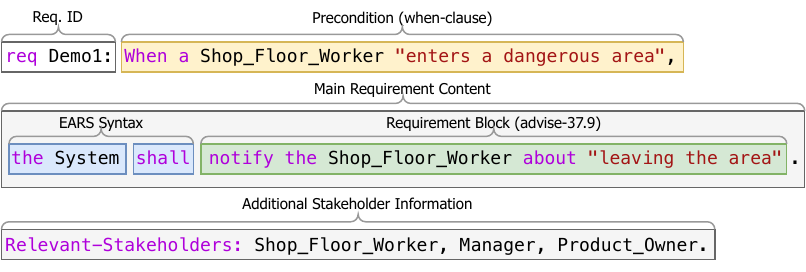}
    \vspace{-1.5em}
    \caption{Example of a Human Monitoring Requirement in \hmreqcnl.}
    \label{fig:grammar-example}
    \Description{Example of a Human Monitoring Requirement following our \hmreq Grammar.}
    \vspace{-1.5em}
\end{figure}

To further aid in detecting such potential value conflicts, we introduce a \emph{Value Conflict Score}, based on the two-dimensional SSA~\cite{schwartz_universals_1992}, which maps correlations among 56 universal values.
Within their configuration, values positioned in close spatial proximity are complementary, while diametrically opposed values demonstrate conflicts~\cite{schwartz_universals_1992,whittlecase2021}.
With this in mind, we calculate the Euclidean distance between all pairs of values and normalize them to get a pairwise potential conflict score between 0 (no conflict) and 1 (maximum conflict). 

\begin{figure}[b!]
\vspace{-.75em}
    \centering
    \includegraphics[width=.90\linewidth]{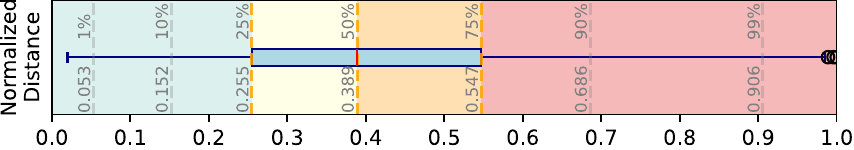}
    \caption{Distribution of possible Value Conflict Scores. Quartiles are highlighted from Green (low) to Red (high), Signalling Potential Conflict Severity.}
    \Description{Distribution of possible Conflict Scores. Quartiles are Highlighted from Green to Red, Signalling Potential Conflict Severity.}
    \label{fig:conflict-score-distribution}
\end{figure}

\begin{figure}[h]
    \centering
    \includegraphics[width=.9\linewidth]{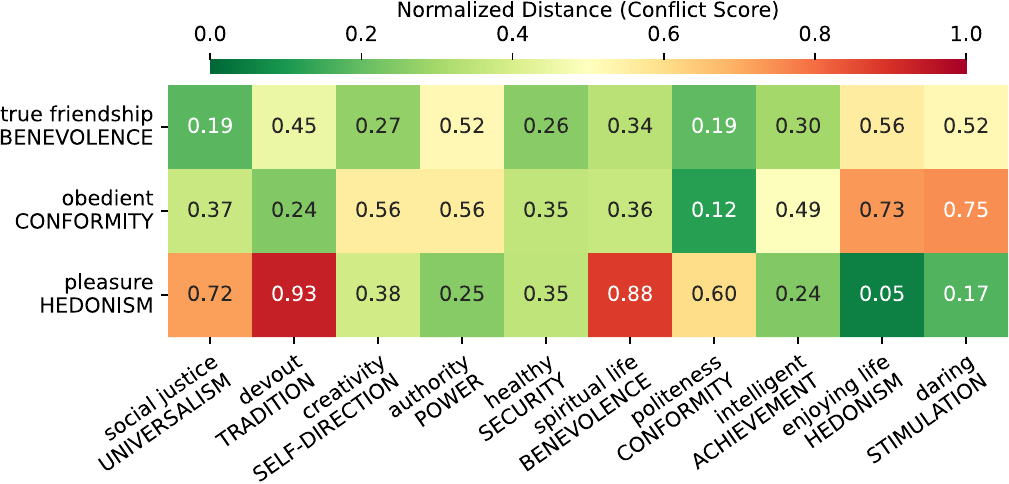}
    \caption{Examples of Potential Value Conflict Score Mappings.}
    \Description{Examples of Potential Conflict Score Mappings.}
    \label{fig:scores-examples}
    \vspace{-.75em}
\end{figure}

We show the overall distribution of possible conflict scores and a selection of example mappings in Figures~\ref{fig:conflict-score-distribution} and~\ref{fig:scores-examples}.
Note that the Value Conflict Score provides a naïve baseline heuristic indication of potential value conflicts; it is not a measure of ethical correctness. It should be used to prioritize value conflicts for stakeholder negotiation, not to automate their resolution.
With this information on potential conflict scores, a requirements engineer can prepare for stakeholder meetings to discuss potential refinements of a requirement (2.3) and come to a final decision on its definition (2.4). 


Revisiting our motivating example, the requirement \emph{\enquote{While a Shop\_Floor\_Worker "is working in dangerous areas", the System shall track "the location" of the Shop\_Floor\_Worker by means of "a GPS sensor". Relevant-Stakeholders: Shop\_Floor\_Worker, Manager, Product\_Owner.}}, may be associated with several stakeholder values and corresponding value statements.
A \textit{Shop Floor Worker} may choose a value related to their privacy, such as \textit{freedom} (\textit{self-direction}) with the value statement \emph{\enquote{I do not want to be identified while being tracked.}}. 
A \textit{Manager}, however, may choose an opposing value related to their \textit{authority} (\textit{power}) with the value statement \emph{\enquote{The system will allow me to find accountable individuals in case of an accident.}}.
Finally, the \textit{Product Owner} may choose a value related to \textit{healthiness}, i.e., protection, of their workers (\textit{security}).
Based on these values, our calculation would result in a potential value conflict score of 0.55 between the \textit{Shop Floor Worker} and the \textit{Manager}. This score is in the top 25\% of highest potential conflict scores, suggesting a high chance of a potential value conflict (cf.~\citefig{conflict-score-distribution}).
The value conflict between the \textit{Manager} and the \textit{Product Owner} is $\approx$0.27, which means a lower chance of conflict between these stakeholders on this specific requirement.
These insights can then aid in stakeholder discussions related to value conflict resolutions and prioritization~\cite{ieee_computer_society_ieee_2021}.
Note that low value conflict scores should not be interpreted as evidence that no conflicts exist. Furthermore, high value conflict scores do not imply that a requirement is unacceptable, but indicate that the involved stakeholder values are distant according to Schwartz’s~\cite{schwartz_universals_1992} value taxonomy and therefore warrant closer examination.



\begin{figure*}[ht]
    \centering
        \includegraphics[width=1\textwidth]{images/tool/value-tool-small.pdf} 
    \caption{Overview Page and Detail View for Requirement R6, Including Potential Value Conflict Highlights.}
    \Description{Screenshot of the tool's overview page showing highlighted value conflicts.}
    \label{fig:tool-overview}
    \vspace{-0.7em}
\end{figure*}

\subsection{Prototype Tool Support}
To support requirements engineers and stakeholders in specifying \hmrs using the \hmreqcnl and aid them in uncovering potential conflicts, we developed two prototype tools: (1) a Visual Studio Code (VSC) extension including a CLI to define \hmrs using \hmreqcnl, and (2) the \hmreqdash, which allows assigning stakeholder values to requirements while automatically calculating potential value conflict scores.

To define our \hmreqcnl, we used Eclipse \emph{Langium}~\cite{_eclipselangium_2021}, an open-source language engineering tool.
Langium enables deployment of a defined grammar as a VSC extension, providing syntax highlighting and basic autocompletion.
Additionally, we used Langium to create a basic command-line interface for the \hmreqcnl to export requirements to \texttt{JSON}, e.g., for further processing in external applications.
Additionally, we added basic language validation checks to ensure that only valid \hmreqcnl requirements are exported. 

To incorporate human values and assist stakeholders in finding potential conflicts, we built a proof-of-concept tool (\hmreqdash) that allows association of human values based on the value taxonomy of Schwartz~\textcite{schwartz_universals_1992} to each relevant stakeholder of a given requirement.
The tool allows importing \hmrs defined with the \hmreqcnl which are then listed on an overview page (cf.~\citefig{tool-overview}).
Each requirement can then be augmented with the human values~\cite{schwartz_universals_1992} and value statements~\cite{pfister_valuecomplemented_2025} of each respective stakeholder. 
After two or more stakeholders have defined their respective human values for a given requirement, the tool calculates the conflict scores of each stakeholder-value pair (cf.~\citesec{value-assignment-conflict-resolution}). 
The average of these value conflict scores determines the intensity of red highlighting on the requirements overview page.
If a user wants to gain more information on the potential value conflicts of a requirement, they can view them in a detail view (cf.~\citefig{tool-overview}).
There, each respective stakeholder-value pair is shown together with its individual potential conflict score.

\section{Defining the CNL Structure}
\label{sec:method-cnl}

To create our \hmreqcnl and identify core verbs and structures, we followed a structured process inspired by the work of Veizaga et al.~\textcite{veizaga_systematically_2021} 
where they developed a CNL specialized in defining functional requirements within the financial domain. In the following, we outline the steps for creating the \hmreqcnl: First, (1) we collected a list of system requirements from requirements datasets (cf. Table~\ref{tab:datasets} DS-1--5) containing potential \hmrs, then, (2) we identified requirements from the dataset that are human monitoring related, and (3) split the selected \hmrs into a training and testing set (for subsequent evaluation -- cf.~\citesec{rq1}).
For the training set, we (4) conducted a thorough analysis and analysed the verbs used within the requirements and mapped them to VerbNet codes and potential word senses.
Once the mapping was complete, we (5) manually selected relevant verbs and senses, and (6) defined the grammar based on the sentence structure found in VerbNet and the respective requirements containing the applicable verb within the dataset.

\subsection{Collecting Requirements (Steps 1-3)}
\label{sec:reqcollection}
We first extracted requirements from five publicly available datasets (cf. Table~\ref{tab:datasets} DS-1--5). 
The datasets include requirements from different domains, ranging from Smart Living Environment Systems for the elderly~\cite{ACTIVAGE_dataset} to Health Care Reporting and Evaluation Systems~\cite{VHCURES_dataset}. 
Some of the extracted requirements follow common patterns, such as EARS~\cite{mavin_easy_2009}, while others are specified in unstructured natural language. 

After consolidation of the datasets and removal of duplicates, we obtained a list of 1596 requirements. 
The first and second author of the paper then checked all requirements and manually marked them as \hmrs if they fell into our definition (cf. Section~\ref{sec:introduction}). For example, the requirement \enquote{Detect when the shower is used} is classified as a \hmr, since it requires some form of monitoring, whereas \enquote{The ratings shall be from a scale of 1-10} requires no human monitoring, and is thus not classified as a \hmr. 
This procedure resulted in a total of 117 \hmrs that we considered for further analysis. We selected a stratified random sample, in which each dataset is a separate stratum, with an 80\% training and 20\% testing split. 
This resulted in a total of 92 \hmrs for developing our \hmreqcnl (training set) and 25 \hmrs for evaluating our CNL (testing set).
The respective amounts of requirements considered for training and testing can be seen in Table~\ref{tab:datasets}. 
Since our selection of \hmrs is not evenly distributed across each dataset, our approach may lead to sampling bias, which we address in Section~\ref{sec:threats}. 

\subsection{Analysing HMRs (Steps 4 and 5)}

We applied automated natural language processing using NLTK~\cite{bird2009natural} to the \hmr training set. For the analysis, we used the wordnet2022 and verbnet3 corpora~\cite{nltk_nltk_data_}.
We first extracted verb lemmas from each requirement and retrieved their corresponding VerbNet classes and associated WordNet senses that start with the given lemma. In cases where NLTK was unable to retrieve a VerbNet class for a lemma, we added that lemma, including its potential synonyms found in WordNet, to a list of auxiliary verbs for manual inspection.

This process resulted in 177 different VerbNet classes found in our training set. In 18 cases, no applicable WordNet sense could be associated with their respective lemma-VerbNet class pair. For 16 lemmas, we were unable to associate a verb lemma to a VerbNet code. 
In a second step, two authors of the paper manually reviewed the individual VerbNet classes, along with their associated verb lemmas and their requirements contexts, to select VerbNet class-lemma pairs they deemed relevant for the human monitoring context. 

Out of the initially collected 177 VerbNet classes, we selected 59 deemed relevant for human monitoring, which we further reduced to 49 through consolidation. 
In cases where no VerbNet code was assigned to the verb lemma, we looked at the verbs' synonyms retrieved from WordNet and assigned a VerbNet class based on an applicable synonym. 
If no synonym was deemed applicable (e.g., the verb \emph{flag} as in \emph{marking a specific situation}), we specified an applicable synonym with a corresponding VerbNet class. 

\subsection{Producing the Grammar (Step 6)}
For each VerbNet code in our refined selection, we leveraged the syntax definitions found in VerbNet to create the basic structure of each grammar rule.
When a lemma appeared in multiple VerbNet classes, we selected only the class most relevant to the domain-specific \hmrs, resulting in a set of 49 different rules based on their VerbNet classes. 

After defining the initial version of our CNL, we attempted to specify all \hmrs from the training set using our CNL. 
Following our Design Science research method~\cite{hevnerDesignScienceInformation2004}, we iteratively refined the grammar rules whenever a requirement could not be expressed with the existing grammar rules and vocabulary. This ensured that the grammar could capture the respective requirement.
As a result, we added several lemmas to existing VerbNet-class-based grammar rules, such as \emph{track} in \texttt{assessment-34.1}. 

Further, we added custom rules for the lemmas \emph{enable} and \emph{subject (to)} as these do not appear in any existing VerbNet class. In these cases, we based the syntactic structure on the respective requirement structure included in the training set.
Following this process, our \hmreqcnl includes 51 requirement-block rules based on VerbNet classes with a total of 87 verbs.
We provide a full list of verbs, the grammar, and the training and testing datasets in the supplementary material.

\section{Evaluation}\label{sec:results}
In the following, we outline our three research questions and  evaluation methodology.

\begin{table*}[t]
\small
\centering
\caption{Datasets used for collecting HMRs and for manual testing of \hmreq. \emph{Monitoring Req.} shows the total of selected requirements within training and testing splits. \emph{Definability} shows if test-set requirements were definable using \hmreqcnl (green: definable, orange: partially definable, red: undefinable). The width of the bars represent relative percentage within a test-set.}
\label{tab:datasets}
\vspace{-5pt}
\begin{tabularx}{\textwidth}{@{}Xcccccl@{}}
\toprule
Dataset & Source & \shortstack{Viewed\\ Req.} & \shortstack{Monitoring\\ Req.} & {Training} & Testing & Definability of Testing Set \\
\midrule
DS-1: ACTIVAGE            & \cite{ACTIVAGE_dataset}          & 310  & 63  & 50  & 13 & \includegraphics[]{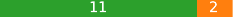} \\
DS-2: MobSTr              & \cite{MobSTr_dataset}            & 69   & 15  & 12  & 3 & \includegraphics[]{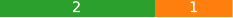}\\
DS-3: NFR\_EXP (PROMISE)  & \cite{PROMISE_exp_dataset}       & 968  & 15  & 12  & 3 & \includegraphics[]{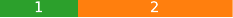}\\
DS-4: VHCURES             & \cite{VHCURES_dataset}           & 180  & 17  & 13 & 4 & \includegraphics[]{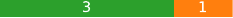}\\
DS-5: WHO SRH 21.4 (NFR)  & \cite{who_dataset}               & 69   & 7   & 5 & 2 & \includegraphics[]{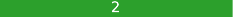} \\
\midrule
DS-6: PURE-subsets        & \cite{ferrari_pure_2017a} via \cite{sahu_reqnet_2025}       &     &    &   &  & \\
\quad DS-6.1: \textit{nenios}                   & \cite{ferrari_pure_2017a} & 48   & 1   & -  & 1 & \includegraphics[]{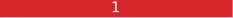} \\
\quad DS-6.2: \textit{phin}                     & \cite{ferrari_pure_2017a} & 129  & 24  & -  & 24 & \includegraphics[]{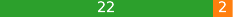} \\
\quad DS-6.3: \textit{ConnectedVehiclePilotNYC} & \cite{ferrari_pure_2017a} & 395  & 11  & -  & 11 & \includegraphics[]{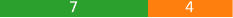} \\
\quad DS-6.4: \textit{automated-insulin-pump}   & \cite{ferrari_pure_2017a} & 15   & 2   & -  & 2 & \includegraphics[]{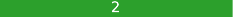}\\
\quad DS-6.5: \textit{EHR-SystemFuncReq LA-DHS} & \cite{ferrari_pure_2017a} & 590  & 18  & -  & 18 & \includegraphics[]{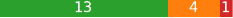} \\
DS-7: QuRE                & \cite{femmer_description_2025}   & 2187 & 20  & -  & 20 & \includegraphics[]{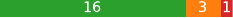}\\
DS-8: WorldVista          & \cite{malik_data_2023a}          & 147  & 8   & -  & 8 & \includegraphics[]{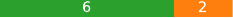} \\
\midrule
DS-9: Dronology HM-Generated    & Original: \cite{cleland-huang_dronology_2018} & - & 20 & -  & 20 & \includegraphics[]{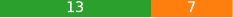} \\
\midrule
Total & & 5107 & 221 & 92 & 129 & \includegraphics[]{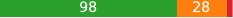} \\
\bottomrule
\end{tabularx}
\vspace{-1.5em}
\end{table*}

\textbf{RQ1: What concepts and structures are required in a CNL to adequately capture human monitoring requirements?} 
With the first research question, we aim to establish the necessary grammar rules for our language, and the general concepts needed to accurately depict \hmrs. This includes selecting an overall requirement template (such as EARS \cite{mavin_easy_2009}) and commonly used verbs for capturing \hmrs.
Additionally, to support our validation, we developed proof-of-concept tools enabling stakeholders to (1) specify requirements in an editor and (2) assign human values to each requirement to aid requirements engineers in detecting value conflicts in stakeholder meetings (cf.~\citesec{rq1})\vspace{0.3em}.

\textbf{RQ2: Can the HM-Req CNL be used to represent requirements of a real CPS?}
With the second research question, we evaluate whether our CNL is capable of capturing requirements of a previously unseen set of \hmrs, more specifically, the Dronology~\cite{cleland-huang_dronology_2018} dataset (cf.~\citesec{rq2})\vspace{0.3em}.

\textbf{RQ3: How do domain experts perceive the usefulness of our CNL and Dashboard?}
Finally, with the third research question, in addition to evaluating our \hmreqcnl with testing datasets, we conducted a survey (n=17) and an additional in-depth interview with a domain expert.
With this, we aim to gain feedback on the perceived usefulness and ease of use of our CNL and proof-of-concept value conflict detection tool \hmreqdash (cf.~\citesec{rq3}). 


\subsection{RQ1: Required CNL Concepts}\label{sec:rq1}
To answer RQ1, we compiled an \hmr dataset from five sources (cf. Table~\ref{tab:datasets} DS-1--5) and divided it into a training and testing split (cf.~\citesec{reqcollection}).
In a first step, we developed our \hmreqcnl based on the training set and subsequently evaluated whether it could accurately capture the requirements of the testing set.
Then, we collected several more human monitoring requirements datasets (cf. Table~\ref{tab:datasets} DS-6--8) that were not utilized during the process of creating our \hmreqcnl. We use these additional requirements to independently validate our \hmreqcnl.

In total, our first testing set comprises 25 requirements (cf. Table~\ref{tab:datasets}). Out of these 25 requirements, we were able to fully specify 19 (76\%) with our \hmreqcnl.
In some cases, a requirement had to be divided into multiple requirements, reducing the complexity of the original unstructured requirement, while preserving its meaning.
The remaining 6 (24\%) requirements could only be captured \textit{partially}, meaning that the requirement could be written in the language, but the grammar did not fully support (1) embedded actor representations, (2) duration statements, or (3) the original semantics.
For example (1), requirement \texttt{FR-7.2.3} \enquote{The system shall be able to identify `pedestrians, cars, [...]'} can only represent the \textit{Pedestrian} actor as a string, but not as a symbol in the language.
Further (2), our CNL lacks support for duration statements such as \emph{\enquote{during the night}}. For example, requirement \texttt{ISE\_Rq\_15} \emph{\enquote{Count the number of steps during the day of the user.}} had to be captured as \emph{\enquote{The System shall track `the number of steps' of the Patient \textbf{every `single' day}.}}.
Lastly (3), the semantics of some requirements were slightly changed. Requirement \texttt{VHCURES-5}, \enquote{The vendor shall be subject to all terms [...]} had to be altered to \enquote{The Vendor shall \textbf{ensure} compliance with all terms [...]'}.

Our second and fully independent (from creation of the \hmreqcnl) testing set includes 84 requirements.
Similar to our initial testing set, we were able to fully specify 66 ($\approx$78.6\%) of all requirements with our \hmreqcnl without issue.
In contrast to our initial testing set, 3 requirements ($\approx$3.6\%) were undefinable using our grammar.
In these cases, our grammar lacked support for specific language constructs, such as an \emph{and} connection in a \emph{When} structure (cf. req. qure498). 
The remaining 15 ($\approx$17.9\%) requirements could be partially captured, with the most common cause being unsupported embedded actor representation\footnote{Percentages have been rounded to the nearest tenth. As a result, the totals may not sum exactly to 100\%.}.

Answering RQ1, $\approx$96.4\% of all testing set requirements were capturable with our \hmreqcnl; $\approx$78.6\% were captured without any issue,
$\approx$17.9\% were partly captured, and $\approx$3.6\% could not be captured without refinements to the grammar.
With minor revisions to the CNL grammar, the \hmreqcnl would be able to capture the remaining requirements, suggesting that our approach resulted in the development of a CNL that successfully captures the human monitoring requirements found in the testing sets. 


\subsection{RQ2: HM-Req CNL Usage}
\label{sec:rq2}
To answer RQ2, we searched for a dataset that contained requirements for a system within the CPS domain. 
The Dronology dataset~\cite{cleland-huang_dronology_2018} provides several publicly available datasets for a drone mission planning and execution system developed at the University of Notre Dame. 
In our case, we selected the ``Requirements Dataset (5/23/2018)''~\cite{drono_website}, which contains 99 natural language requirements, as well as several design definitions, tasks, and trace links to source code. 

Since the initial Dronology dataset did not contain explicit \hmrs, we generated requirements  using Google's Gemini 2.5 Pro LLM~\cite{comanici_gemini_2025}. We prompted the LLM with (1) the instruction to generate 20 \hmrs, (2) our definition of \hmrs, (3) the general EARS syntax template, (4) information on the Dronology system, and (5) the available system requirements of Dronology. 
The resulting requirements were initially validated by the first author for general correctness. To confirm that the generated requirements were, in fact, valid and relevant for a drone system, we asked two domain experts who had experience using Dronology to assess all requirements and adapt any that would not be valid in this context. 
This resulted in 10 requirements that could be accepted without edits, 4 requirements that could be accepted with minor edits, and 6 requirements that had to be rejected. 
To compensate the rejections, we manually added 6 new requirements with validation of a domain expert.

Overall, we were able to capture \emph{all} generated Dronology requirements using our \hmreqcnl. 65\% were definable without issue, whereas 35\% were partially definable (cf. Table~\ref{tab:datasets}). Similar to the evaluation of our testing sets (cf. Section~\ref{sec:rq1}), the majority of partial representation came from unsupported actor embeddings. Additional iterative refinements to the \hmreqcnl grammar would further improve the results. Nevertheless, actors were correctly representable within free-text blocks of the grammar. 

Answering RQ2, our results suggest that our \hmreqcnl is capable of correctly capturing human monitoring requirements from real-world CPS.

\subsection{RQ3: Perceived Usefulness of the CNL and Dashboard}
\label{sec:rq3}
To gain first insights on the usefulness of both our \hmreqcnl and the \hmreqdash, we conducted an exploratory anonymous survey (n=17) with questions based on a subset of the technology acceptance model~\cite{davis1989technology}.
The survey comprised 15 questions organized into three sections: (1) demographic profiling, (2) usefulness and ease-of-use assessment during \hmr specification tasks using the \hmreqcnl, and (3) usefulness assessment of the value conflict detection prototype \hmreqdash.
Questions on the perceived usefulness and ease-of-use employed a 7-point Likert scale (labelled: \emph{extremely unlikely} to \emph{extremely likely} and \emph{fully disagree} to \emph{fully agree}, respectively). Both sections (2) and (3) contained optional open-ended feedback questions.

\textbf{Setup/Tasks:} In part 2, \emph{specification of \hmrs using our \hmreqcnl}, participants were asked to define two \hmrs based on unstructured text using our CNL in an online VS Code instance, and then rated their perceived usefulness and ease-of-use. 
In part 3, participants viewed screenshots and an online version of the \emph{\hmreqdash} and then evaluated its perceived usefulness.\\

\subsubsection{Demographic Results}
In total, 17 participants completed our survey. Participants represented several professional backgrounds, including 8 researchers, 3 requirements engineers, 3 software engineers, and 3 in hybrid roles (2 SE/research; 1 SE/RE/research).
We recruited participants through targeted email invitations sent to researchers and practitioners in the field of CPS, RE, and Value-Based Engineering. They were further encouraged to forward the study to potentially interested colleagues.
The professional experience was distributed nearly evenly (0-2 years:~3; 3-5 years:~5; 6-10 years:~5; >10 years:~4). 
Further, most participants (15 out of 17) have had previous experience with requirements engineering, and over 60\% had prior experience with CPS. 
All but one participant were familiar with User Stories and Use Cases, 8 participants were familiar with unstructured natural language requirements, and 2 participants were familiar with EARS.
The demographic data suggests that the participants had sufficient experience with requirements engineering to evaluate our tools. However, their exposure to structured requirement approaches (e.g., EARS) was limited.\\

\subsubsection{Usefulness and Ease of Use of HM-Req CNL}
Participants considered our framework's approach of using the \hmreqcnl to specify \hmrs as likely to be useful (slightly likely: 4, quite likely: 11, extremely likely: 2). 
Regarding the ease of specifying \hmrs, four participants considered using the CNL extremely likely to be beneficial, while nine and four participants rated it as quite likely and slightly likely, respectively.
14 participants found the CNL to be at least quite likely to prevent misunderstandings between different stakeholders compared to requirements written in unrestricted natural language.
All participants stated that the CNL would improve \hmr quality, with responses ranging from extremely likely (5) to slightly likely (6).
Feedback regarding usability was positive, with 16 out of 17 participants agreeing (10) or fully agreeing (6) that learning our CNL would be easy for them.
A similar consensus was formed regarding the ease of use of our Visual Studio Code extension, with all but one participant being in slight or more agreement that using the CNL with our extension and defining \hmrs is easy to use. 

Additionally, we asked if the predefined structure of the \hmreqcnl, constraining the specification of requirements, guided them toward writing better, more precise requirements.
A common sentiment of participants was that while the restrictions were inconvenient when trying out the grammar in Visual Studio Code, they were confident that with more practice, our CNL has the potential to improve the quality of their requirements. 
Looking at the \hmr definition attempts of participants, we observed that several participants had trouble using our CNL correctly (i.e., using free-form text instead of verbs after the modal verb \textit{shall}, missing punctuation) while others were able to correctly redefine the requirements. We expect users to gain more experience after using our CNL for some time, and providing tutorials.
One participant mentioned that the recommendations from the extension helped them to follow the EARS template better without making mistakes. Another participant thought it was helpful for non-native English speakers to follow correct grammar rules.\\

\subsubsection{Usefulness of our Proof-of-Concept Value Conflict Tool}
All but one participant found our \hmreqdash potentially useful, with 12 participants finding the Dashboard quite useful or better. 
Similarly, 14 participants specified it is quite or extremely likely to facilitate more effective stakeholder discussions and negotiations.
Further, 16 participants found it quite or extremely likely that the \hmreqdash would make it easier to detect potential stakeholder value conflicts.
Regarding the definition of stakeholder values, 15 participants found the \hmreqdash to be slightly likely to be useful or better, one stating it to be slightly unlikely, and one neither. 

Related to where the tool would fit in the participant's workflow, most (10) stated in free-text that the tool would be best suited during initial stakeholder workshops. 
Specifically, participants stated that the tool could be used before going into detailed negotiation, \enquote{to catch certain disagreements beforehand}. 
However, two participants suggested that it may be too soon to use in early stakeholder workshops, since requirements evolve rapidly in the early design phases. They would prefer using the tool in the detailed system design stages. 
It was also suggested to be used in verification and validation phases (2), to \enquote{catch conflicts before going into development}.\\

\subsubsection{Interview}
In addition to the survey, we conducted a semi-structured interview with a domain expert in CPS (a researcher experienced in software engineering, human-machine interaction, and CPS application development) to obtain detailed feedback on our proof-of-concept \hmreqcnl and \hmreqdash. 
Before the interview, the participant participated in our survey. The goal was to gain additional in-depth insights on the approach of defining \hmrs using our \hmreqcnl compared to their existing RE methods. 

The expert's current approach for defining requirements mainly used unstructured text, collected through informal interviews, in GitHub issues. 
They stated that they liked the approach of having to follow a structure during requirements definition: \emph{\enquote{[...] if we are following the format, then the assumption is that the requirements will be much clearer for everyone else in the team or for all stakeholders to better understand [that requirement].}}
Further, they suggested a two-step system, where natural language requirements would be translated into the \hmreqcnl and later manually adjusted within an editor. 
When asked what they would add to the language, they highlighted the need to clearly specify the feedback a human receives when interacting with humans: \emph{\enquote{[...] for example, as a drone pilot, I want to schedule a go-to-waypoint command and expect that the drone sends me an acknowledgement so that I know that drone is going to do that task and not something else.}}
Regarding how our potential value conflict tool would fit in their process, they stated that, while they had not previously considered it, they thought our \hmreq framework \emph{\enquote{is something we need in terms of development. There was no formal analysis of how implementing these features are going to create these ethical conflicts or privacy issues.}}.




Answering RQ3, our results indicate strong participant agreement on the usefulness of our approach using our CNL and the \hmreqdash. We are therefore convinced that continuing to improve our framework based on the feedback will be valuable for researchers and CPS engineers.

\section{Discussion}
\label{sec:discussion}

Based on the evaluation of our framework, this section discusses key findings, limitations, and promising directions for future work to enhance the HM-Req framework.

\textbf{Extended Grammar:}
Our evaluation confirmed that while our grammar can capture a vast majority of requirements, there is still potential for improvement.
As mentioned in Sections~\ref{sec:rq1} and~\ref{sec:rq2}, we found that our CNL cannot capture all \hmrs within our testing sets. A common occurrence was the limitation regarding requirements using timing-related statements (e.g., \textit{during the night}), which we plan to add to the grammar in the next iteration.
Participants also suggested specific additions (e.g., evaluate, authenticate) and removals (e.g., allow, change) from the grammar's verb lexicon.
Further, one participant was concerned about the \hmreqcnl allowing free text literals to restrict the future possibilities related to automated analysis. 
They suggested eliminating free text literals and adding the ability to define custom concepts in the language, replacing the current free text literals. While we do believe that having some leniency through free text literals is beneficial in requirements elicitation, we agree that our grammar currently allows for excessive usage of them. We intend to address this by refining the EARS pre-statements of our grammar (cf. Listing~\ref{lst:requirement_structure}) to conform to more specific grammar rules instead of using string literals. 

Beyond direct syntax, feedback proposes expanding the CNL's conceptual scope. 
Our interview partner suggested incorporating the notion of \emph{feedback} a human receives during human-machine interactions. 
While this can be modelled with additional requirements in the current grammar, we recognize the value of making it a first-class concept.
Participants also suggested that values should be included within the grammar to be cross-checked immediately while writing the requirement, enabling earlier detection of value conflicts and thus reduced resolution cost. 
We deliberately excluded value specification from the initial version of the \hmreqcnl to reduce its overall complexity, but it remains a compelling direction for future development. 


\textbf{Extended Tool Support:}
Some participants found that existing tooling was inconvenient to use, especially regarding in-line completion of our Visual Studio Code extension. This limitation stems from the current implementation, which relies on basic by-name completion (e.g., for Stakeholders, Actors, and other keywords) and lacks full abstract syntax tree (AST) awareness. An AST-based approach would enable context-sensitive suggestions and a more guided writing experience.
Further suggestions included implementing clearer error messages and warnings in case the specified requirement does not follow the CNL as well as integrations with other IDEs.
We plan to improve our tool to provide a more robust and intuitive user experience in the future, lowering the barrier to adoption for users of the CNL.

\textbf{Extended Negotiation/Conflict Resolution Tool Support: }
Our survey participants provided valuable insights into potential future additions regarding the proof-of-concept \hmreqdash.
One suggestion was to expand the scope of our value analysis beyond a single requirement. 
Our current tool only detects potential value conflicts between stakeholders within one requirement, whereas a future version could support inter-requirement conflict detection. This would allow requirements engineers to identify how the values associated with one requirement might conflict with another, thus providing a clearer picture of the overall value landscape. This may be especially beneficial when working with requirements that focus on one stakeholder at a time.
Another participant suggested better traceability and management of identified conflicts within the tool. They highlighted the need for functionality to formally document the resolution process, for example, by explicitly \textit{accepting} a value conflict and recording the reasoning for doing so. This would create a clear audit trail for design decisions and enhance accountability.


\textbf{LLM-supported Requirements and Value Identification:}
A promising topic for future work is the integration of LLMs into our \hmr framework. 
Both survey participants and our interview partner highlighted the potential of integrating LLMs, for example, to automatically convert unstructured requirements to \hmreqcnl requirements.
We believe that an LLM, when instructed to formulate unstructured requirements based on the syntax tree of our grammar, may be a valuable tool for helping to define \hmreqcnl requirements.
LLMs also have the potential to aid requirements engineers in the manual identification of \hmrs from an existing set of system requirements by providing a confidence score on how likely a given requirement is human monitoring related.
Further, LLMs could be asked to generate potential values and value statements based on stakeholder user profiles or personas, which could then be imported into \hmreqdash for potential conflict assessment. Using LLMs for these tasks would alleviate some of the manual effort currently required using the \hmreq framework.

\section{Threats to Validity}\label{sec:threats}

Our work is subject to several validity threats. While we have shown that our CNL is applicable to real-world requirements, the limited number of use cases may result in other domains/types of systems not yet being fully covered. 
A potential threat to \emph{external validity} is the risk of sampling bias. Our training set of 92 \hmrs was compiled from five datasets from different domains. However, the distribution of these requirements was not uniform across the source datasets. We attempted to mitigate this by using stratified sampling to ensure all source datasets were represented. Nevertheless, the imbalance may result in our \hmreqcnl being inadvertently optimized for the vocabulary and structures of the more dominant datasets. 
However, our results show that our \hmreqcnl can successfully define \hmrs from several previously unseen datasets (cf. Sections~\ref{sec:rq1} and~\ref{sec:rq2}), suggesting generalizability beyond the dataset it was trained on. Our manual selection of \hmrs for training and testing is subject to potential \emph{selection bias}. To mitigate this, the first author initially identified a candidate set of HMRs from the original datasets, which was then independently reviewed and validated by a senior researcher. Discrepancies were resolved through iterative discussions until a consensus was reached.

The limited number of participants in the survey and interview poses a potential threat to the \emph{statistical conclusion validity} of our study, as the sample size does not provide sufficient quantitative evidence for broad generalization.
Therefore, the feedback, while valuable, should be seen as indicative rather than a conclusive, generalizable measure of our HM-Req \emph{CNL's} and \mbox{HM-Req} \emph{Dashboard's} overall acceptance. 
To further assess the usefulness and usability of our \hmreqcnl and \hmreqdash, we intend to conduct a large-scale user study with a controlled experiment. 
The calculation of our naïve baseline potential Value Conflict Score poses a potential threat to the \emph{construct validity} of our study. Since we did not have access to the underlying value data Schwartz used for their smallest space analysis, we had to resort to calculating the Euclidean distances between each value. Naturally, this introduces measurement imprecision and may lead to scores not being a true reflection of the underlying taxonomy.
Additional studies in real-world projects are required to validate the effectiveness of the score in detecting value conflicts.

\section{Related Work}
\label{sec:relwork}


\textbf{Approaches and DSLs for Requirements Specification: }
Several approaches and languages have been proposed for specifying different types of requirements. 
Early work by Fuchs and Schwitter~\textcite{fuchs_attempto_1995} developed the Attempto CNL, which allows specifying system requirements that can then be translated into Prolog clauses.
Veizaga et al.~\textcite{veizaga_systematically_2021} defined a process to systematically build a CNL for requirements. Based on that process, they developed Rimay, a CNL to define functional requirements in the financial domain, which also provided the basis for our approach when developing the \hmreqcnl.
Further, Gomez-Vazquez and Cabot~\cite{gomez2025towards} introduced MERLAN, a DSL for specifying requirements of multimodal interfaces for AI-enhanced systems. 
Similarly, SEMKIS~\cite{jahic2023semkis}, a DSL designed to support engineers in specifying requirements for recognizing skills of neural networks. Researchers at NASA built a restricted natural language called FRETISH~\cite{giannakopoulou_formal_2020}, which enables the definition of system requirements. In addition, they provided tooling support, including semantic hints and sentence structure colouring. 
While these approaches allow definition of structured software requirements using CNLs or DSLs, none of them focus on the specific domain of human monitoring in CPS.

\textbf{Human Values \& Value Conflicts: }
Whittle et al.~\textcite{whittlecase2021} explored how human values can be incorporated in the software-development process of real-world projects. They use Schwartz's value taxonomy~\cite{schwartz_universals_1992} as a starting point to define relevant stakeholder values as so-called value portraits. 
They argue that values capture the \emph{why} of requirements engineering as an addition to the regular \emph{what} (functional requirements) and \emph{how} (non-functional requirements).
In our work, we use value-statements based on the idea of value-portraits to add context information as to \emph{why} a stakeholder has associated a specific value to a requirement.
Perera et al.~\cite{perera_continual_2020} leveraged Schwartz's taxonomy~\cite{schwartz_universals_1992} in their Continual Value(s) Assessment (CVA) framework, which utilises goal models to describe relationships of requirements to specific stakeholder values. The relationships can either be positive or negative, highlighting potential value tradeoffs and conflicts during development. They continuously refine values during system development, leading to a more complete picture of potential values. 
Distinct from their work, we leverage a CNL to define domain-specific \hmrs aiming to reduce ambiguity and vagueness during requirements definition. Further, after associating stakeholder values to requirements, we provide a Value Conflict Score that highlights potential conflicts that need to be resolved during stakeholder workshops.
Winter et al.~\textcite{winter_advancing_2019} introduced \textit{Value Q-Sort}, a mixed method combining semi-structured interviews with a sorting problem.
Each statement from interviews was associated with Schwartz's value taxonomy~\cite{schwartz_universals_1992}. 
Wohlrab et al.~\textcite{wohlrab_supporting_2024} introduced the concept of \textit{value tactics}, which are design decisions that suggest mechanisms to consider based on Schwartz~\cite{schwartz_universals_1992} values and sub-values.
Besides academic literature, the standard for addressing ethical concerns during system design~\cite{ieee_computer_society_ieee_2021} describes a detailed process to design systems considering individual and societal ethical values. Instead of using a value taxonomy like Schwartz~\cite{schwartz_universals_1992}, they suggest identifying and prioritizing values based on the ethical theories \emph{utilitarian ethics}, \emph{virtue ethics}, and \emph{duty ethics}.
Existing research provides methods for eliciting human values of stakeholders and integrating them into the software engineering process. However, they do not directly link \hmrs to stakeholder values nor provide a quantifiable way to highlight potential value conflicts for stakeholder discussions.



\textbf{Human Interaction/Collaboration and Monitoring: }
Adams~\textcite{adams_human-robot_2005} made the case for the need to incorporate machines in existing human team activities, such as search and rescue missions in contaminated areas. They suggest that before designing the system (i.e., specifying its requirements), one needs to understand the current human team's activities. 
Agrawal et al.~\textcite{agrawal_next_2020} conducted interviews with firefighters to gain insights on how a human-machine collaborative emergency response system could be improved by designing the system for situational awareness~\cite{endsley_designing_2016}. 
They defined Use Cases in unstructured language based on initial requirements discussions, but did not consider the human values of participants during elicitation.
Calinescu et al.~\textcite{calinescu_maintaining_2021} proposed an approach using a MAPE control loop in combination with sensor arrays to collect a human's biometric data to improve driver attentiveness in shared-control autonomous driving. They detail methods to collect data using sensors such as gaze position and heart rate, but do not specify formal \hmrs. 
Current literature suggests that research in human-machine interaction rarely highlights human values or ethics conflicts of stakeholders that may arise during human monitoring.

\section{Conclusion}\label{sec:conclusion}
In this paper, we addressed the challenge of defining unambiguous, value-aware human monitoring requirements in CPS.
We introduced the \hmreq framework, which combines (1) a CNL for formally specifying human monitoring requirements with (2) the systematic association of stakeholder values based on Schwartz's taxonomy~\cite{schwartz_universals_1992} with human monitoring requirements and surfacing potential conflicts via a Value Conflict Score.
Our HM-Req framework is supported by a Visual Studio Code extension, providing basic syntax highlighting and auto-completion for our \hmreqcnl, and a proof-of-concept \hmreqdash, allowing stakeholder value assignment and calculation of Value Conflict Scores.

We evaluated our framework through a mixed-methods evaluation: Results indicate that our \hmreqcnl is able to capture human monitoring requirements of real-world CPS and several requirements datasets. During evaluation, \hmreqcnl successfully captured over 96\% ($\approx$78.6\% fully, $\approx$17.9\% partial) of human monitoring requirements from real-world datasets. Furthermore, an exploratory survey and expert interview suggest that practitioners perceive our \hmreqcnl and \hmreqdash as useful for both specifying requirements and enabling value-aware stakeholder discussions.

Ultimately, our work proposes a framework to specify structured human monitoring requirements, making value conflicts explicit and quantifiable during system design, aiming to elevate the consideration of human values during human-machine interaction from an abstract ideal into a systematic engineering practice.

\vspace{.5em}
\noindent\textbf{Data availability:} We provide all supplementary material and additional exemplar descriptions as part of our repository:~\url{https://github.com/Ethical-Human-Machine-Interaction/hmreq}.

\normalsize
\small

\bibliography{cleaned}

\end{document}